# Fractional differential and fractional integral modified-Bloch equations for PFG anomalous diffusion and their general solutions

Guoxing Lin*

Carlson School of Chemistry and Biochemistry, Clark University, Worcester, MA 01610, USA

**Abstract**

Anomalous diffusion widely exists in many systems such as polymeric or biological systems. However, the studying of anomalous diffusion by pulsed field gradient (PFG) diffusion technique still faces challenges due to the complexity of PFG anomalous diffusion theory. Two different research groups have successfully obtained consistent theoretical results for space-fractional diffusion by employing the modified Bloch equation method. However, for time-fractional diffusion and general fractional diffusion, the modified Bloch equations reported by these two groups have different forms and, therefore, yield inconsistent results; additionally, one group's reported results only consider the first period attenuation, while another group's reported results cannot reduce to short gradient pulse (SGP) approximation result. The discrepancy in these reported modified Bloch equations may arise from different ways of combining the fractional diffusion equation with the precession equation where the time derivatives have different derivative orders and forms, which makes the combination complicated. Moreover, to the best of my knowledge, the general PFG signal attenuation expression including finite gradient pulse width (FGPW) effect for time-space fractional diffusion based on the fractional derivative has yet to be reported by other methods. In practical applications, the FGPW effect may not be neglectable, particularly in MRI. Thus, it is necessary to find different combination strategies to build new Bloch equations and to obtain general PFG signal attenuation expressions for anomalous diffusions. Here, based on different combination strategy, two new modified Bloch equations are proposed, which belong to two significantly different types: a differential type based on the fractal derivative and an integral type based on the fractional derivative. The merit of the integral type modified Bloch equation is that the original properties of the contributions from linear or nonlinear processes remain unchanged at the instant of the combination. The general solutions including the FGPW effect were derived from these two equations as well as from two other methods: a method observing the signal intensity at the origin and the recently reported effective phase shift diffusion equation method. The obtained PFG signal attenuation expressions can be reduced to the SGP approximation expressions. The relaxation effect was also considered. It is found that the relaxation behavior influenced by fractional diffusion based on the fractional derivative deviates from that of normal diffusion. The general solution agrees perfectly with continuous-time random walk (CTRW) simulations as well as reported literature results. The new modified Bloch equations is a valuable tool to describe PFG anomalous diffusion in NMR and MRI.

## I. INTRODUCTION

Pulsed field gradient (PFG) [1,2,3,4,5] techniques have been an important tool to study diffusion. There are many theoretical frames that have been developed to successfully explain PFG normal diffusion. However, for anomalous diffusion which has been found in many systems such as in polymer or biological systems [6 7, 8, 9, 10], the PFG theory still faces challenges. In PFG experiments, for theسake of convenience, anomalous diffusion may be classified into the following three types: general fractional diffusion $\{0<\alpha,\beta\leq 2\}$, space-fractional diffusion $\{\alpha=1, 0<\beta\leq 2\}$, and time-fractional diffusion $\{0<\alpha\leq 2,\beta=2\}$, where $\alpha$ and $\beta$ are the time and space derivative parameters respectively [11,12,13,14 15, 16] (the $\alpha$ is used for time derivative with respect to earlier PFG anomalous diffusion reports in NMR). Unlike normal diffusion, the anomalous diffusion [17,18,19] has a mean square displacement that is not linearly proportional to diffusion time and its probability distribution function is also non-Gaussian. These distinct characteristics make it complicated to analyze PFG anomalous diffusion.

There are many theoretical efforts to investigate the PFG anomalous diffusion [20, 21, 22,23,24, 25, 26, 27, 28, 29, 30]. One of the most important PFG theoretical methods is the modified-Bloch equation method, which has been very successful in normal diffusion such as the familiar Torrey-modified Bloch equation [31,32]. Two different groups have successfully obtained consistent PFG signal attenuation expressions for space-fractional diffusion by employing the modified Bloch equation method [14,33], however, for time-fractional diffusion and general fractional diffusion, their proposed modified Bloch equations have different forms and, therefore, yield inconsistent theoretical results. Additionally, in the PGSE or PGSTE experiments as shown in Fig. 1, the time $t$ can be divided into three periods: $0<t\leq\delta$, $\delta<t\leq\Delta$, $\Delta<t\leq\Delta+\delta$; for time-fractional and general fractional diffusion. The results from the reported modified Bloch equation proposed by one of these two groups [14] can not be reduced to the short gradient pulse (SGP) approximation's result obtained by [15,16], while another group's [33] result did not give the PFG signal attenuation for all three periods in a typical PFG experiment. The SGP approximation method [4,5] is a basic theoretical method that can give PFG signal attenuation for an ideal experimental situation. The experiment employs ideal gradient pulses whose intensity $g$ can be assumed to be infinitely strong (which is just for theoretical purposes rather than technical purposes). To get a certain value of $\gamma g\delta$, the required gradient pulse width $\delta$ for such a strong gradient will be infinitely narrow, and the diffusion can be neglected inside each gradient pulse. The ideal dephasing gradient pulse creates a space modulated phase structure $\exp(-i\gamma g\delta z)$ inside the sample where $z$ is the position, which will be refocused by the rephasing pulse. The signal attenuation results from only the diffusion during the diffusion delay $\Delta$. For a spin starting from the origin (for a homogeneous system, the attenuation is the same everywhere), the signal attenuation can be described as $S(\Delta)=S(0)\int_{-\infty}^{\infty}P(z,\Delta)\exp(-iz\cdot\gamma g\delta)dz$ [4,5] where $P(z,\Delta)$ is the spin particle probability distribution function at time $\Delta$ and position $z$, $S(0)$ and $S(t)$ are the signal intensities at the beginning of the first dephasing gradient pulse and at the end of the rephasing pulse, respectively. The obtained SGP approximation result is $E_{\alpha,1}\left(-D_f(t)K_{SGP}^{\beta}t^{\alpha}\right)$ [15,16] for time-space fractional diffusion based on the fractional derivative, where $E_{\alpha,1}(x)=\sum_{n=0}^{\infty}\frac{x^n}{\Gamma(1+n\alpha)}$ is Mittag-Leffler function [6,11]. This method greatly simplifies the analyzing of PFG diffusion by neglecting the diffusion inside the dephasing and rephasing gradient pulses. The other theoretical results usually should agree with the SGP results. The discrepancy in these reported results may arise from the combination of diffusion and precession. The combination is complex since precession has a time derivative $\partial/\partial t$ while the fractional diffusion has a significantly different time derivative with the derivative order $0 < \alpha \leq 2$. The modified Bloch equation method has been a fundamental and useful method for normal diffusion. It is necessary to build a modified Bloch equation for fractional diffusion that can account for the three types of fractional diffusion and its solution can be reduced to the SGP result. Such a modified Bloch equation could be built using different combination strategies.

This paper is an attempt to build the desired modified Bloch equations for fractional diffusion and to obtain general PFG signal attenuation expressions that are essential to analyzing PFG anomalous diffusion experiments. To achieve this goal, two theoretical treatments were proposed: an observing the signal intensity at the origin method (or observing only the signal intensity method), and new modified Bloch equations for fractional diffusion. Two different types of fractional modified Bloch equations were built: one is a differential equation type, and the other is an integral equation type. The integral type Bloch equation is different from the conventional differential type of modified Bloch equation [32]. The integral type equation can combine the contributions from different processes such as diffusion, precession and relaxation into an integral with the same time increasement $dt$, which overcomes the difficulties mentioned above due to the different time derivatives in the fractional derivative model. Two different derivative models, fractal derivative [34,35] and fractional derivative [11,12,13,19] model, were used to build these equations. The observing only signal intensity method and the new modified Bloch equations yield the same or equivalent signal attenuation equations for both free normal and anomalous diffusions. It was also derived in this paper that these equations for free anomalous diffusion can be obtained by employing the reported effective phase shift diffusion equation (EPSDE) method [15]. The general solution for PFG fractional diffusion was derived and the results include finite gradient pulse width (FGPW) effect [4,5] and it can be reduced to the SGP result. The results agree with reported results from other methods in the literature. The results also agree perfectly with continuous-time random walk (CTRW) [36] simulations performed in this paper. The relaxation effect in PFG anomalous diffusion experiments was also considered, and it was found that, unlike the normal diffusion and fractional diffusion based on the fractal derivative, the relaxation behavior affected by fractional diffusion based on fractional derivative model deviates from the common expression $\exp(-t/T_2)$ [4,5,9]. The general PFG signal attenuation formalisms of free anomalous diffusion and the new modified Bloch equations will be valuable tools in PFG NMR and MRI.

## II. DIFFERENT SIMULTANEOUS PROCESSES IN PFG DIFFUSION EXPERIMENTS

In PFG experiments, the magnetization evolution is affected by simultaneous processes such as Lamour precession, diffusion and relaxation. For simplicity, one-dimensional diffusion will be studied here. The spin magnetization can be described as [4,5]

$$M_{xy}(z,t) = M_x(z,t) + iM_y(z,t), \tag{1}$$

where $z$ is the coordinate vector. The magnetic field can be described as $\boldsymbol{B}(z,t) = \boldsymbol{B}_0 + \boldsymbol{g}(t) \cdot z$, where $\boldsymbol{B}_0$ is the exterior magnetic field, and $\boldsymbol{g}(t)$ is the time-dependent gradient. In a rotating frame rotating around magnetic field at angular frequency $\omega = -\gamma \mathbf{B}_0$ where $\gamma$ is the gyromagnetic ratio, the spin precession can be described by [4,5]

$$\frac{\partial}{\partial t} M_{xy}(z,t) = -i\gamma \mathbf{g}(t) \cdot z M_{xy}(z,t). \qquad (2)$$

The relaxation of the transverse component of magnetization can be described as [4,5,9]

$$\frac{\partial}{\partial t} M_{xy}(z,t) = -\frac{M_{xy}(z,t)}{T_2}, \qquad (3)$$

where $T_2$ is the spin-lattice relaxation time constant.

The normal diffusion can be described by [37]

$$\frac{\partial}{\partial t} M_{xy}(z,t) = D \frac{\partial^2}{\partial z^2} M_{xy}(z,t), \qquad (4)$$

where $D$ is the diffusion coefficient. The modified Bloch equation for normal diffusion can be obtained by the combination of Eqs. (2)-(4), which is [36]

$$\frac{\partial}{\partial t} M_{xy}(z,t) = D \frac{\partial^2}{\partial z^2} M_{xy}(z,t) - i\gamma \mathbf{g}(t) \cdot z M_{xy}(z,t) - \frac{M_{xy}(z,t)}{T_2}. \qquad (5)$$

The same time derivative $\partial/\partial t$ in Eqs. (2)-(4) is one of the premises for the straightforward combination of these equations. However, compared to $\partial/\partial t$, the time fractional derivative of fractional diffusion has a different derivative order $\alpha$, $0 < \alpha \leq 2$, and a different form, thus, it may not be appropriate to directly combine the fractional diffusion equation with the precession equation or relaxation equation. For instance, the time-space fractional diffusion could be modeled by the fractal derivative as [34,35]

$$\frac{\partial}{\partial t^\alpha} M_{xy}(z,t) = D_{f_1} \frac{\partial}{\partial z^\beta} M_{xy}(z,t), \qquad (6)$$

where $D_{f_1}$ is the fractional diffusion coefficient for the fractal derivative model with units $m^\beta/s^\alpha$, $\frac{\partial}{\partial t^\alpha}$ and $\frac{\partial}{\partial z^\beta}$ are the time and space fractal derivative defined in Appendix A.

The time-space fractional diffusion can also be modeled by the fractional derivative as

$${}_t D_*^\alpha M_{xy}(z,t) = D_{f_2} \frac{\partial^\beta}{\partial |z|^\beta} M_{xy}(z,t), \qquad (7)$$

where $D_{f_2}$ is the fractional diffusion coefficient for the fractional derivative model with units $m^\beta/s^\alpha$, $\frac{\partial^\beta}{\partial |z|^\beta}$ is the space fractional derivative defined in Appendix B, and ${}_t D_*^\alpha$ is the Caputo fractional derivative

defined as [11-13]

$$_tD_*^\alpha f(t) := \begin{cases} \dfrac{1}{\Gamma(m-\alpha)} \int_0^t \dfrac{f^{(m)}(\tau)d\tau}{(t-\tau)^{\alpha+1-m}}, & m-1 < \alpha < m, \\ \dfrac{d^m}{dt^m} f(t), & \alpha = m. \end{cases} \qquad (8)$$

Both the fractal time derivative $\dfrac{\partial}{\partial t^\alpha}$ and the Caputo time derivative $_tD_*^\alpha$ are distinct from $\dfrac{\partial}{\partial t}$. For space-fractional diffusion with $\{\alpha=1, 0<\beta\leq 2\}$, the time fractional derivative reduces to $\dfrac{\partial}{\partial t}$. There is no issue with the combination of the diffusion and procession, and the results from the modified Bloch equations proposed by different research groups agree with each other. However, for time fractional diffusion $\{0<\alpha\leq 2, \beta=2\}$ and general fractional diffusion $\{0<\alpha, \beta\leq 2\}$, when $\alpha \neq 1$, these reported results disagree with each other. The difference between the time derivatives $_tD_*^\alpha$ and $\dfrac{\partial}{\partial t}$ makes it difficult to combine of these different processes. Before we figure out how to combine these motions to build new modified Bloch equations in section [V], let us first look at the effective phase shift diffusion equation method reported in [15] in next section, and the observing the signal intensity at the origin method in section [IV]. The PFG same signal attenuation expressions can be obtained from both methods.

## III. GENERAL SIGNAL ATTENUATION EQUATIONS BASED ON THE EFFECTIVE PHASE SHIFT DIFFUSION METHOD

The EPSDE method describes non-refocused phase distribution due to the coeffect of the precession and diffusion with a phase diffusion equation built in the phase space. The EPSDE method has given an exact PFG signal attenuation expression $S(t) = \exp\left[-\int_0^t DK^\beta(t')dt'^\alpha\right]$ for fractional diffusion based on the fractal derivative model. However, for the fractional derivative model, the EPSDE method has only given $E_{\alpha,1}\left[-D_f(t)K_{SGP}^\beta t^\alpha\right]$, which is an SGP approximation result. In this section, the general PFG signal attenuation expressions for both fractal and fractional derivative models are derived in another way. These signal attenuation expressions is solved from the $q$-space accumulating phase diffusion equations obtained by the Fourier transform of the effective phase shift diffusion equations. The $q$-space equation employed here is different from the conventional $k$-space diffusion equation which is obtained by the Fourier transform of real space diffusion equation.

For simplicity, a one-dimensional diffusion along $z$ direction is considered. In PFG experiments, as it is mentioned in the previous section, field gradient pulses create a time and position dependent magnetic field $\boldsymbol{B}(z,t) = \boldsymbol{B}_0 + \boldsymbol{g}(t)\cdot z$ [4,5]. The spin moment precesses about the magnetic field with Lamour frequency $\omega = -\gamma \boldsymbol{B}(z,t)$. In a rotating frame rotating around a magnetic field at an angular frequency $\omega_0 = -\gamma \boldsymbol{B}_0$, the phase $\psi(z,t)$ of non-diffusing spins at $z$ is

$$\psi(z,t) = -\boldsymbol{K}(t)\cdot z, \qquad (9)$$

where

$$\boldsymbol{K}(t) = \int_0^t \gamma \boldsymbol{g}(t)dt' \qquad (10a)$$

is the wavenumber vector. For a simple PGSE experiment with constant gradient as shown in Fig. 1, the wavenumber is [5]

$$\mathbf{K}(t) = \begin{cases} \gamma \mathbf{g} t, 0 < t \leq \delta \\ \gamma \mathbf{g} \delta, \delta < t \leq \Delta \\ \gamma \mathbf{g}(\Delta + \delta - t), \Delta < t \leq \Delta + \delta \end{cases}. \qquad (10b)$$

The gradient pulses create a space modulated magnetization in the non-diffusing spin system, which can be described as [5]

$$M_{xy}(z,t) = S(t)\exp(-i\mathbf{K}(t)\cdot z), \qquad (11)$$

where $S(t)$ is the signal intensity (it is often called amplitude, but both positive and negative $S(t)$ could be created in the same sample if selective pulses were used, so $S(t)$ is called signal intensity instead of amplitude here). The typical PFG experiments consist of two gradient pulses. One is the dephasing gradient pulse, and the other is the rephasing gradient pulse. The rephasing gradient pulse counteracts the effect of the dephasing gradient pulse, which makes the wavenumber vector $\mathbf{K}(t)$ return to 0 at the end of the rephasing pulse. As a result, the magnetization of non-diffusing spins in the whole sample will be refocused. Therefore, there is no signal attenuation in PFG experiments for non-diffusing spins.

While spins in a diffusion process will have a time-dependent position in the sample, the accumulating phase of a diffusing spin is [7]

$$\phi(t) = -\int_0^t \gamma \mathbf{g}(t') \cdot z(t')dt', \qquad (12)$$

where $z(t')$ is the displacement vector. $\phi(t)$ is the net accumulating phase due to diffusion in PFG experiments at time $t$. $\phi(t)$ is different from $\psi(z,t)$ defined by Eq. (9). $\psi(z,t)$ is the space modulated phase determined by wavenumber and is refocused at the end of the rephasing pulse. The values of $\phi(t)$ depend on the coeffect of the $\mathbf{g}(t')$ set in the experiment and the $z(t')$ determined by the diffusion path, which is in the range of $-\infty < \phi < \infty$ rather than $-\pi < \phi < \pi$. Since superdiffusion can be theoretically possible to have an infinitely large $\phi^2(t)$, one may confuse how NMR can measure a superdiffusion. It is not an issue to measure superdiffusion by NMR. The $\phi(t)$ value only affects the value of $\cos(\phi(t))$, which is the projection factor of the magnetization at the x-axis. $-1 \leq \cos(\phi(t)) \leq 1$ is always true regardless if $\phi(t)$ is infinitely large or small. The fast diffusion usually corresponds fast attenuation of the NMR signal. Therefore, PFG NMR can measure subdiffusion as well as superdiffusion. The PFG signal attenuation can be obtained by averaging over all possible phases by [5,7]

$$S(t) = S(0)\int_{-\infty}^{\infty} P(\phi,t)\exp(+i\phi)d\phi, \qquad (13)$$

where $P(\phi,t)$ is the accumulating phase probability distribution function, and, $S(0)$ and $S(t)$ are the signal intensity at the beginning of the first dephasing gradient pulse and the signal intensity at the end of the rephasing pulse, respectively. Getting $P(\phi,t)$ is essential to calculating PFG signal attenuation based on Eq. (13). The value of $S(0)$ will set as 1 and it will be dropped out throughout this paper (namely, the signal intensity $S(t)$ is normalized). Conventionally, the $P(\phi,t)$ is assumed to be a Gaussian function, which is the familiar Gaussian phase distribution (GPD) approximation. However, the solutions from the effective phase shift diffusion equations show that the free normal diffusion in a homogeneous sample has an exact Gaussian phase distribution, while the anomalous diffusions have non-Gaussian phase distributions [15].

In the EFSDE method reported in [15], the self-diffusion process for normal or anomalous diffusion is described as a sequence of independent random jumps with waiting times $\Delta t_1$, $\Delta t_2$, $\Delta t_3$,..., $\Delta t_n$, and a corresponding sequence of jumps with displacement lengths $\Delta z_1$, $\Delta z_2$, $\Delta z_3$,…, $\Delta z_n$. Eq. (12) can be rewritten as [15]

$$\begin{aligned}\phi(t) &= -\sum_{i=1}^{n}\gamma\Delta t_i \boldsymbol{g}(t_i)\cdot\left(\sum_{m=1}^{i}\Delta z_m + z_0\right) \\ &= -\gamma\sum_{m=1}^{n}\left[\sum_{i=1}^{n}\Delta t_i \boldsymbol{g}(t_i) - \sum_{i=1}^{m-1}\Delta t_i \boldsymbol{g}(t_i)\right]\cdot\Delta z_m - \gamma\sum_{i=1}^{n}\Delta t_i \boldsymbol{g}(t_i)\cdot z_0 , \\ &= -\sum_{m=1}^{n}[\boldsymbol{K}(t) - \boldsymbol{K}(t_{m-1})]\cdot\Delta z_m - \boldsymbol{K}(t)\cdot z_0\end{aligned} \quad (14)$$

where $\boldsymbol{K}(t)$ is the wavenumber defined in Eq. (10), and the terms $\Delta t_i \boldsymbol{g}(t_i)$ and $\Delta z_m$ are assumed to be two non-correlated terms. For most cases in PFG experiments, $\boldsymbol{K}(t_{tot}) = 0$ where $t_{tot}$ is the time at the end of the rephasing gradient pulse, while $\boldsymbol{K}(t_{tot}) \neq 0$ can be found in certain experiments [38]. Only $\boldsymbol{K}(t_{tot}) = 0$ will be considered in Eq. (14) and we have [15]

$$\phi(t_{tot}) = -\sum_{m}\boldsymbol{K}(t_m)\cdot\Delta z_m, K(t_{tot}) = 0, \quad (15)$$

where $\boldsymbol{K}(t_m) \approx \boldsymbol{K}(t_{m-1})$ is used as $\Delta\boldsymbol{K}(t_m)\cdot\Delta z_m$, is neglectable and $\Delta t_m$ can be small. The $-\sum_{m}\boldsymbol{K}(t_m)\cdot\Delta z_m$ term in Eq. (15) can be viewed as an effective random jump with a jump displacement $-\boldsymbol{K}(t_m)\cdot\Delta z_m$ and a jump waiting time $\Delta t_m$. Such a random walk is a phase diffusion which can be gotten from the corresponding spin diffusion in the real space by scaling the jump length $\Delta z(t)$ with a scaling factor $-K(t)$. Therefore, the effective phase diffusion coefficient is [15] $D_{\phi eff}(t) = K^{\beta}(t)D$, noted usually $K(t) \geq 0$. By replacing coordinate $z$ with $\phi$, and $D_f$ with $D_{\Phi eff}(t)$ respectively from the one-dimensional real space diffusion equations, Eqs. (4), (6) and (7), the following effective phase diffusion equations can be obtained [15]:

$$\frac{\partial}{\partial t}P(\phi,t) = K^2(t)D\frac{\partial^2}{\partial z^2}P(\phi,t), normal\ diffusion \quad , \quad (16\text{ a})$$

$$\frac{\partial}{\partial t^{\alpha}}P(\phi,t) = K^{\beta}(t)D_{f_1}\frac{\partial}{\partial z^{\beta}}P(\phi,t), fractal\ derivative \quad , \quad (16\text{b})$$

and

$$_tD_*^{\alpha}P(\phi,t) = K^{\beta}(t)D_{f_2}\frac{\partial^{\beta}}{\partial |z|^{\beta}}P(\phi,t), fractional\ derivative \quad . \quad (16\text{c})$$

By performing Fourier transform $\hat{F}f(x) = \int_{-\infty}^{\infty} f(x)\exp(+iqx)dx$ on both sides of Eqs. (16a)-(16c), as

$$\hat{F}\left(\frac{\partial^2}{\partial z^2}f(z);q\right) = -q^2 f(q), \quad \hat{F}\left(\frac{\partial^\beta}{\partial z^\beta}f(z);q\right) = -q^\beta f(q) \text{ [34,35], and } \hat{F}\left(\frac{\partial^\beta}{\partial |z|^\beta}f(z);q\right) = -q^\beta f(q) \text{ [11-12]}$$

where $q$ is wavenumber, we get the following equations:

$$\frac{\partial}{\partial t}P(q,t) = -K^2(t)q^2 D P(q,t), normal\ diffusion, \qquad (17a)$$

$$\frac{\partial}{\partial t^\alpha}P(q,t) = -K^\beta(t)q^\beta D_{f_1} P(q,t), fractal\ derivative, \qquad (17b)$$

and

$$_tD_*^\alpha P(q,t) = -K^\beta(t)q^\beta D_{f_2} P(q,t), fractional\ derivative, \qquad (17c)$$

Here, $q$ is dedicated to the Fourier transform of the effective phase equation, which must not be confused with the field gradient pulse induced wavenumber $K(t)$ defined by Eq. (10). In a PFG experiment, the PFG signal attenuation can be obtained by [4,5,7]

$$S(t) = \int_{-\infty}^{\infty} P(\phi,t)\exp(+i\phi)d\phi = p(1,t). \qquad (18)$$

Eq. (18) indicates that the PFG signal attenuation is equal to $P(q,t)$ with $q = 1$. Substituting Eq. (18) and $q = 1$ into Eq. (17a), we get

$$\frac{\partial}{\partial t}S(t) = -K^2(t)DS(t). \qquad (19)$$

The solution of Eq. (19) is $S(t) = \exp\left(-\int_0^t DK^2(t')dt'\right)$, which reproduces the reported theoretical results [39]. For a simple PGSE or PGSTE experiment, the integration of $\exp\left(-\int_0^t DK^2(t')dt'\right)$ yields the familiar expression $\exp\left[-D\gamma^2 g^2 \delta^2 (\Delta - \delta/3)\right]$ for a free normal diffusion [4,5]. For the fractional diffusion based on the fractal derivative model, by substituting Eq. (18) and $q = 1$ into Eq. (17b), we get

$$\frac{\partial}{\partial t^\alpha}S(t) = -K^\beta(t)D_{f_1}S(t). \qquad (20)$$

From Eq. (20), the PFG signal attenuation expression can be obtained as

$$S(t) = \exp\left(-\int_0^t D_{f_1}K^\beta(t')dt'^\alpha\right), \qquad (21)$$

where $dt'^\alpha = \alpha t'^{\alpha-1}dt'$. Eq. (21) reproduces the reported results by the EPSDE method [15] and ISA method [16]. While for the fractional derivative model, by substituting Eq. (18) and $q = 1$ into Eq. (17c), we get

$$_tD_*^\alpha S(t) = -K^\beta(t)D_{f_2}S(t). \qquad (22)$$

The Eq. (22) will be solved in section [V] as its same or equivalent equations will be obtained by observing the signal intensity at the origin method in section [IV] and by the proposed fractional integral modified Bloch equation in section [V].

## IV. OBSERVING THE SIGNAL INTENSITY AT THE ORIGIN

The Eqs. (19), (20) and (22) for PFG signal attenuation can be derived in an alternative way. In a homogeneous sample, the signal intensity of magnetization is the same everywhere in the sample. As discussed in the previous section, the gradient pulses result in space-modulated magnetization $M_{xy}(z,t) = S(t)\exp(-i\mathbf{K}(t)\cdot z)$ described by Eq. (11) in a non-diffusing spin system. For a diffusion spin system, the possibilities of spins at an arbitrary position $z$ moving toward opposite directions $z \pm \Delta z$ are equal inside a homogenous sample. The opposite motions yield an averaging phase $\exp(-i\mathbf{K}(t)\cdot(z+\Delta z)) + \exp(-i\mathbf{K}(t)\cdot(z+\Delta z)) = \cos(\mathbf{K}(t)\cdot\Delta z)\exp(-i\mathbf{K}(t)\cdot z)$ that only affects the signal intensity $S(t)$ but does not change the phase $\exp(-i\mathbf{K}(t)\cdot z)$. The phase of a diffusion spin system is still modulated by the precession like that in the non-diffusing spin system, and it is still described by Eq. (11), namely $M_{xy}(z,t) = S(t)\exp(-i\mathbf{K}(t)\cdot z)$. The different diffusion types only determine what type the function of $S(t)$ belongs but they do not affect the average phase. Therefore, $M_{xy}(z,t) = S(t)\exp(-i\mathbf{K}(t)\cdot z)$ holds for both free normal and anomalous diffusion in a homogeneous sample. By substituting Eq. (11) into Eq. (4), we have

$$\frac{\partial}{\partial t}[S(t)\exp(-i\mathbf{K}(t)\cdot z)] = D\frac{\partial^2}{\partial z^2}[S(t)\exp(-i\mathbf{K}(t)\cdot z)]. \tag{23}$$

From Eq. (23), since $\nabla^2 \exp(-i\mathbf{K}(t)\cdot z) = -\mathbf{K}^2(t)\exp(-i\mathbf{K}(t)\cdot z)$, it is ready to get

$$\frac{\partial}{\partial t}[S(t)\exp(-i\mathbf{K}(t)\cdot z)] = -D\mathbf{K}^2(t)[S(t)\exp(-i\mathbf{K}(t)\cdot z)]. \tag{24}$$

At $z = 0$, $\exp(-i\mathbf{K}(t)\cdot z) = 1$ and $S(t)\exp(-i\mathbf{K}(t)\cdot z) = S(t)$, if one observes only the signal intensity at the origin, the Eq. (24) becomes

$$\frac{\partial}{\partial t}S(t) = -DK^2(t)S(t). \tag{25}$$

Eq. (25) is the same as Eq. (19). The solution for Eq. (25) is $S(t) = \exp\left[-\int_0^t DK^2(t')dt'\right]$, which has been obtained in [39*]. The above observing signal intensity at the origin method can be further understood in the following context: for a homogeneous sample, if one has limited technique sources allowing him to monitor only the time-dependent signal intensity $S(t)$ but not the phase, the observer will not be aware of the phase change as $|\exp(-i\mathbf{K}(t)\cdot z)| = 1$ at any locations. Eq. (24) should still govern his observed signal intensity data, but it will be in a reduced form that is the same as Eq. (25) obtained by observing the signal intensity at the origin method.

Again, by substituting Eq. (11), $M_{xy}(z,t) = S(t)\exp(-i\mathbf{K}(t)\cdot z)$ to the fractional diffusion equation, Eq. (6) based on the fractal derivative, and applying $\frac{\partial}{\partial z^\beta}\exp(-i\mathbf{K}(t)\cdot z) = -|K(t)|^\beta \exp(-i\mathbf{K}(t)\cdot z)$, we have

$$\frac{\partial S(t)\exp(-i\mathbf{K}(t)\cdot z)}{\partial t^\alpha} = -|K(t)|^\beta \exp(-i\mathbf{K}(t)\cdot z) \qquad (26a)$$

Similarly, if we only observe the signal intensity at $z = 0$, $\exp(-i\mathbf{K}(t)\cdot z) = 1$ and $S(t)\exp(-i\mathbf{K}(t)\cdot z) = S(t)$, we get

$$\frac{\partial S(t)}{\partial t^\alpha} = -D_{f_1} K^\beta(t) S(t). \qquad (26b)$$

In most cases, $K(t) \geq 0$, the absolute value symbol $|\ |$ will be dropped out in the rest of this paper. Eq. (26b) is the same as Eq. (20) and its solution is Eq. (21), $S(t) = \exp\left(-\int_0^t DK^\beta(t')dt'^\alpha\right)$, which is the signal intensity evolution equation based on the fractal derivative model.

By substituting $M_{xy}(z,t) = S(t)\exp(-i\mathbf{K}(t)\cdot z)$ into the fractional diffusion equation Eq. (7) based on the fractional derivative model [***] and employing $\frac{\partial^\beta}{\partial |z|^\beta}\exp(-i\mathbf{K}(t)\cdot z) = -K^\beta(t)\exp(-i\mathbf{K}(t)\cdot z)$, we have

$$_tD_*^\alpha[S(t)\exp(-i\mathbf{K}(t)\cdot z)] = -D_{f_2}(t)K^\beta(t)[S(t)\exp(-i\mathbf{K}(t)\cdot z)] \quad . \qquad (27)$$

Similarly, in a homogeneous sample, by observing the signal intensity at $z = 0$, $\exp(-i\mathbf{K}(t)\cdot z) = 1$ and $S(t)\exp(-i\mathbf{K}(t)\cdot z) = S(t)$, we get

$$_tD_*^\alpha S(t) = -D_{f_2}(t)K^\beta(t)S(t). \qquad (28)$$

Eq. (28 reproduces Eq. (22) that is obtained from the previous section, and its solution will be given in section [V.B.2].

The above method is applicable in a homogeneous spin system. Since the NMR signal amplitude is homogeneously ubiquitous, the result would be the same if the signal amplitude is observed from the origin or any other location. In the following, two modified Bloch equations will be built and their general solution for free anomalous diffusion will be given.

## V. MODIFIED BLOCH EQUATIONS AND THEIR GENERAL SOLUTIONS

Conventionally, the modified Bloch equation was built by combining the precession equation and the diffusion equation directly into a differential equation. However, as the time derivative of the fractional derivative includes the time integration, besides the ordinary differential type equation for the fractal derivative model, an integral type equation for the fractional derivative model will be built.

### A. Differential type modified Bloch equation by the fractal derivative

The fractional diffusion equation based on the fractal derivative model, Eq. (6), can be rewritten as

$$\frac{\partial M_{xy}(z,t)}{\partial t} = \alpha t^{\alpha-1} D_{f_1} \frac{\partial}{\partial z^\beta} M_{xy}(z,t). \tag{29}$$

By combining Eqs. (29) with Eqs. (2) and (3), the modified Bloch equation based on fractal derivative can be obtained as

$$\frac{\partial M_{xy}(z,t)}{\partial t} = \alpha t^{\alpha-1} D_{f_1} \frac{\partial}{\partial z^\beta} M_{xy}(z,t) - i\gamma \mathbf{g}(t)\cdot z M_{xy}(z,t) - \frac{M_{xy}(z,t)}{T_2}. \tag{30}$$

Substituting $M_{xy}(z,t) = S(t)\exp(-i\mathbf{K}(t)\cdot z)$ into Eq. (30), gives

$$\frac{\partial S(t)}{\partial t} = -\left(\alpha t^{\alpha-1} D_{f_1} K^\beta(t) + \frac{t}{T_2}\right) S(t). \tag{31}$$

The solution of Eq. (31) is

$$S(t) = \exp(-\frac{t}{T_2})\exp\left(-\int_0^t D_{f_1} K^\beta(t') dt'^\alpha\right). \tag{32}$$

If the relaxation effect is neglected, Eq. (32) is equivalent to Eq. (21). Additionally, when $\beta = 2$ and in the first period $0 < t \le \delta$, Eq. (31) is reduced to $\frac{\partial S(t)}{\partial t} = -\left(D_{f_1}\alpha(\gamma g)^2 t^{\alpha+1} + \frac{t}{T_2}\right) S(t)$. That is consistent with the results $\frac{\partial S(t)}{\partial t} = -Dg^2 t^{2-\nu} S(t), \nu = 1-\alpha$, obtained by spectral function and echo damping method reported in [40]. Contrasted to the result in [40], Eq. (31) is more general and it can be applied the whole time, $0 < t \le \Delta + \delta$, and to random gradient pulse shapes in PFG general fractional diffusion experiments.

### B. Fractional integral modified Bloch equation by the fractional derivative

#### 1. Building the fractional integral modified Bloch equation

In the fractional derivative model, the time derivative operator such as Caputo time derivative includes the integral, making it difficult to combine the diffusion equation with precession equation. Nevertheless, the combination can be carried out in a different way. The precession equation, Eq. (2) can be rewritten as

$$\partial M_{xy}(z,t) = -i\gamma \mathbf{g}(t)\cdot z M_{xy}(z,t)\partial \tau. \tag{33}$$

The Caputo fractional derivative has the following property [***]:

$$J^\alpha_t D^\alpha_* \mu(t) = u(t) - \sum_{k=0}^{m-1} u^{(k)}(0^+)\frac{t^k}{k!}, \tag{34}$$

where $J^{\alpha}(f(t)) = \int_0^t \frac{(t-\tau)^{\alpha-1}}{\Gamma(\alpha)} f(\tau) d\tau$ [11-12]. By operating $J^{\alpha}$ on both sides of the fractional diffusion equation, Eq. (7), we get

$$M_{xy}(z,t) = \sum_{k=0}^{m-1} M^{(k)}(z,0^+) \frac{t^k}{k!} + J^{\alpha}\left[D_{f_2} \frac{\partial^{\beta}}{\partial |z|^{\beta}} M_{xy}(z,t)\right], \quad (35a)$$

equivalent to

$$M(z,t) = \sum_{k=0}^{m-1} M^{(k)}(z,0^+) \frac{t^k}{k!} + \int_0^t \frac{(t-\tau)^{\alpha-1}}{\Gamma(\alpha)} D_{f_2} \frac{\partial^{\beta}}{\partial |z|^{\beta}} M_{xy}(z,t) d\tau, \quad (35b)$$

Where $M_{xy}^{(0)}(z,0^+) = M_{xy}(z,0)$ and $M_{xy}^{(1)}(z,0^+) = \frac{\partial}{\partial t} M_{xy}(z,0^+)$. In Eq. (35b), the term $\frac{(t-\tau)^{\alpha-1}}{\Gamma(\alpha)} D_{f_2} \frac{\partial^{\beta}}{\partial |z|^{\beta}} M_{xy}(z,t)$ is the diffusion related attenuation during the interval $d\tau$. During the same interval, the precession alters the magnetization by $-i\gamma \mathbf{g}(t) \cdot z M_{xy}(z,t) \partial \tau$ based on Eq. (33). The above two simultaneous changes in the magnetization can be combined inside the same integral, and we eventually get

$$M_{xy}(z,t) = \sum_{k=0}^{m-1} M_{xy}^{(k)}(z,0^+) \frac{t^k}{k!} + \int_0^t \left\{\left(\frac{(t-\tau)^{\alpha-1}}{\Gamma(\alpha)} D_{f_2} \frac{\partial^{\beta}}{\partial |z|^{\beta}} - i\gamma \mathbf{g}(t) \cdot z\right) M_{xy}(z,\tau)\right\} d\tau. \quad (36),$$

the modified Bloch equation built upon the fractional derivative. Eq. (36) is significantly unlike the modified Bloch equation proposed in references [14,33]. We may set $\frac{\partial}{\partial t} M_{xy}(z,0^+) = 0$, that has been suggested by Mainardi et al. [11,13] to get a continuous transition from $\alpha = 1^-$ to $\alpha = 1^+$. The $\frac{\partial}{\partial t} M_{xy}(z,0^+) = 0$ condition will be utilized in the paper. For a homogeneous sample, the magnetization is described by Eq. (11). By performing partial differential on $M_{xy}(z,\tau) = S(\tau) \exp(-i\mathbf{K}(\tau) \cdot z)$, we can acquire $-i\gamma \mathbf{g}(\tau) \cdot z M_{xy}(z,\tau) \partial \tau = \partial M_{xy}(z,\tau) - \exp(-i\mathbf{K}(\tau) \cdot z) \partial S(\tau)$, which can be substituted into Eq. (36) to yield

$$M_{xy}(z,t) = \sum_{k=0}^{m-1} M^{(k)}(z,0^+) \frac{t^k}{k!} + \int_0^t \left\{\left(\frac{(t-\tau)^{\alpha-1}}{\Gamma(\alpha)} D_f \frac{\partial^{\beta}}{\partial |z|^{\beta}}\right) M_{xy}(z,\tau)\right\} d\tau$$

$$+ \int_0^t \partial [M_{xy}(z,\tau) - \exp(-i\mathbf{K}(\tau) \cdot z) \partial S(\tau)]$$

$$= \int_0^t \left\{\left(\frac{(t-\tau)^{\alpha-1}}{\Gamma(\alpha)} D_f |\mathbf{K}(\tau)|^{\beta}\right) M_{xy}(z,\tau) - \exp(-i\mathbf{K}(\tau) \cdot z) \frac{d}{d\tau} S(\tau)\right\} d\tau + M_{xy}(z,t)$$

which produces

$$-\left(\frac{(t-\tau)^{\alpha-1}}{\Gamma(\alpha)}D_f|\mathbf{K}(\tau)|^{\beta}\right)S(\tau)\exp(-i\mathbf{K}(\tau)\cdot\mathbf{r})-\exp(-i\mathbf{K}(\tau)\cdot\mathbf{r})\frac{d}{d\tau}S(\tau)=0 \quad,$$

$$\frac{d}{d\tau}S(\tau)=-\left(\frac{(t-\tau)^{\alpha-1}}{\Gamma(\alpha)}D_f|\mathbf{K}(\tau)|^{\beta}\right)S(\tau),$$

$$S(t)=\sum_{k=0}^{m-1}S^{(k)}(0^+)\frac{t^k}{k!}-J^{\alpha}\left(D_f|\mathbf{K}(\tau)|^{\beta}S(\tau)\right) \Leftrightarrow {}_tD_*^{\alpha}S(t)=-K^{\beta}(t)D_{f_2}S(t). \quad (37)$$

Eq. (37) is the same as Eqs. (22) and (28) in sections [III] and [IV].

The relaxation Eq. (3) can be written as $\partial M_{xy}(z,\tau)=-\frac{M_{xy}(z,\tau)}{T_2}\partial\tau$ which can be added into the integral in Eq. (36) to give

$$M_{xy}(z,t)=\sum_{k=0}^{m-1}M_{xy}^{(k)}(z,0^+)\frac{t^k}{k!}+\int_0^t\left\{\left(\frac{(t-\tau)^{\alpha-1}}{\Gamma(\alpha)}D_{f_2}\frac{\partial^{\beta}}{\partial|z|^{\beta}}-i\gamma\mathbf{g}(t)\cdot z-\frac{1}{T_2}\right)M_{xy}(z,t)\right\}d\tau, \quad (38),$$

the integral type modified Bloch equation including the $T_2$ relaxation. For the free diffusion in a homogeneous system, Eq. (38) reduces to

$$S(t)=\sum_{k=0}^{m-1}S^{(k)}(0)\frac{t^k}{k!}+\int_0^t\left\{\left(\frac{(t-\tau)^{\alpha-1}}{\Gamma(\alpha)}D_fK^{\beta}(t)-\frac{1}{T_2}\right)S(\tau)\right\}d\tau \quad. \quad (39)$$

The Eqs. (38) and (39) are notably different from the modified Bloch equations for normal diffusion and fractional diffusion based on the fractal derivative. Eqs. (38) and (39) will give a signal attenuation that is not proportional to $\exp(-\frac{t}{T_2})$ which will be discussed further in section [VII]. The solution of Eq. (39) will be given in the next section.

When $\alpha=1, \beta=2$, by operating $\frac{\partial}{\partial t}$ on both sides of Eq. (38), it is easy to get

$$\frac{\partial}{\partial t}M_{xy}(z,t)=D\frac{\partial^2}{\partial z^2}M_{xy}(z,t)-i\gamma\mathbf{g}(t)\cdot zM_{xy}(z,t)-\frac{1}{T_2}M_{xy}(z,t), \quad (40)$$

which reproduces Eq. (5), the modified Bloch equation for normal diffusion [32]. For space fractional diffusion, whose $\alpha=1$, Eq. (38) reduces to

$$\frac{\partial}{\partial t}M_{xy}(z,t)=\left[D_{f_2}\frac{\partial^{\beta}}{\partial|z|^{\beta}}-i\gamma\mathbf{g}(t)\cdot z-\frac{1}{T_2}\right]M_{xy}(z,t). \quad (41),$$

which is consistent with the results reported in [14,33]. In a homogeneous system, Eq. (41) reduces to

$$\frac{\partial}{\partial t}S(t)=\left[-K^{\beta}(t)D_{f_2}-\frac{1}{T_2}\right]S(t), \quad (42)$$

which yields the PFG signal attenuation

$$S(t) = \exp(-\frac{1}{T_2})\exp\left[-D_{f_2}(\gamma g\delta)^\beta (\Delta - \frac{\beta-1}{\beta+1}\delta)\right]. \tag{43}$$

Eq. (43) agrees with the results reported from both modified Bloch equations in Refs. [14,33].

*2. General PFG signal attenuation expression for fractional derivative model*

First, considering the simple free anomalous diffusion without $T_2$ relaxation, we have derived two equivalent PFG signal attenuation equations:

$$\begin{cases} {}_tD_*^\alpha S(t) = a(t)S(t) \\ S(t) = \sum_{k=0}^{m-1} S^{(k)}(0^+)\frac{t^k}{k!} + J^\alpha(a(t)S(t)), m-1 < \alpha < m \end{cases}, \tag{44}$$

where

$$a(t) = -K^\beta(t)D_{f_2} = \begin{cases} -(\gamma g t)^\beta D_{f_2}, 0 < t \leq \delta \\ -(\gamma g \delta)^\beta D_{f_2}, \delta < t \leq \Delta \\ -(\gamma g)^\beta (\Delta + \delta - t)^\beta D_{f_2}, \Delta < t \leq \Delta + \delta \end{cases}. \tag{45}$$

Eq. (44) is for general fractional diffusion with $\{0 < \alpha, \beta \leq 2\}$, includeing time-fractional diffusion, space-fractional diffusion and normal diffusion. The similar type of fractional equation as Eq. (44) has been solved by the Adomian decomposition Method [41,42, 43,44, 45]. According to the results from these references, the solution of Eq. (44) is [44,45]

$$S(t) = \sum_{n=0}^{\infty} S_n(t), \tag{46a}$$

where

$$S_0(t) = \sum_{k=0}^{m-1} S^{(k)}(0^+)\frac{t^k}{k!}, m-1 < \alpha < m, \tag{46b}$$

and

$$\begin{aligned} S_n(t) &= J^\alpha(a(t)S_{n-1}(t)) \\ &= -\int_0^t \frac{(t-\tau)^{\alpha-1}}{\Gamma(\alpha)} D_{f_2} K^\beta(\tau) S_{n-1}(\tau) d\tau \\ &= -\int_0^t \frac{D_{f_2} K^\beta(\tau) S_{n-1}(\tau) d(t-\tau)^\alpha}{\alpha \Gamma(\alpha)} \end{aligned}. \tag{46c}$$

For free diffusion in a homogeneous sample in PGSE or PGSTE experiments as shown in Fig. 1, the time $t$ can be divided into three periods: $0 < t \leq \delta$, $\delta < t \leq \Delta$, $\Delta < t \leq \Delta + \delta$. If the $S^{(1)}(0^+) = 0$ condition is used [11,13], the following can be obtained:

(a) *PFG signal attenuation under SGP approximation*: $\delta$ is neglectable and the diffusion inside each gradient pulse can be neglected. We get $S_0(\Delta) = 1$, $S_n(\Delta) = \dfrac{\left(-D_f(\gamma g \delta)^\beta \Delta^\alpha\right)^n}{\Gamma(1+n\alpha)}$, and

$$S(\Delta) = \sum_{n=0}^{\infty} S_n(t) = E_{\alpha,1}\left(-D_f K_{SGP}^\beta \Delta^\alpha\right), K_{SGP} = \gamma g \delta. \tag{47}$$

Eq. (47) replicates the SGP approximation result obtained via $S(\Delta) = S(0)\int_{-\infty}^{\infty} P(z,\Delta)\exp(-iz\cdot\gamma g\delta)dz$ in references [15,16].

*(b). Single pulse attenuation*: this is an ideal situation, the first gradient pulse is regular, but the second gradient pulse is infinitely narrow which is dedicated to counteracting the effect of the first gradient pulse. Based on the Adomian decomposition method, we get $S_0(t) = 1$, $S_n(t) = \left(-D_f(\gamma g)^\beta t^{(\alpha+\beta)}\right)^n \prod_{k=1}^{n} \frac{\Gamma(1+(k-1)\alpha+n\beta)}{\Gamma(1+k(\alpha+\beta))}$,

$$S(t) = \sum_{n=0}^{\infty} S_n(t) = 1 + \sum_{n=0}^{\infty}\left(-D_f(\gamma g)^\beta t^{(\alpha+\beta)}\right)^n \prod_{k=1}^{n} \frac{\Gamma(1+(k-1)\alpha+n\beta)}{\Gamma(1+k(\alpha+\beta))} \quad (48)$$
$$= E_{\alpha,1+\beta/\alpha,\beta/\alpha}\left(-D_f(\gamma g)^\beta t^{\alpha+\beta}\right)$$

where $E_{\alpha,\eta,\gamma}(x) = \sum_{n=0}^{\infty} c_n x^n, c_0 = 1, c_n = \prod_{k=0}^{n-1} \frac{\Gamma((k\eta+\gamma)\alpha+1)}{\Gamma((k\eta+\gamma+1)\alpha+1)}$ is a Mittag-Leffler type function. [46], which is consistent with the results from the modified Bloch equation proposed by Hanyga et al. in Ref. [33]

*(c). General PFG signal attenuation*: the PGSE or PGSTE experiment as shown in Fig. 1 includes three periods: $0 < t_1 \le \delta$, $\delta < t_2 \le \Delta$, $\Delta < t_3 \le \Delta+\delta$. The integration is tedious, which can be calculated with computer assistance. However, we can get the first and second terms as the following:

$S_0(t) = 1$,

$$S_1(t) = J^\alpha(a(t)S_0(t)) = -\int_0^t \frac{(t-\tau)^{\alpha-1}}{\Gamma(\alpha)} D_{f_2} K^\beta(\tau)d\tau = -\frac{D_{f_2}(\gamma g)^\beta}{\Gamma(1+\alpha)} \times$$

$$\begin{cases} \alpha t^{\alpha+\beta} B(\beta+1,\alpha), & 0 < t \le \delta \\ \alpha t^{\alpha+\beta}\left[B(\beta+1,\alpha) - B\left(\frac{\delta}{t};\beta+1,\alpha\right)\right] - \delta^\beta(t-\delta)^\alpha, & \delta < t \le \Delta \\ \left\{\alpha t^{\alpha+\beta}\left[B(\beta+1,\alpha) - B\left(\frac{\delta}{t};\beta+1,\alpha\right)\right] - \delta^\beta\left[(t-\delta)^\alpha - (t-\Delta)^\alpha\right]\right\} - \\ \int_\Delta^t \alpha(t-\tau)^{\alpha-1}(\Delta+\delta-\tau)^\beta d\tau, & \Delta < t \le \Delta+\delta \end{cases} \quad (49)$$

where B(x,y) and B(a;x,y) are the Beta function and imcomplete Beta function. When $t = \Delta+\delta$, Eq. (49) gives

$$S_1(\Delta+\delta) = J^\alpha(a(t)S_0(t))$$
$$= -\int_0^{\Delta+\delta} \frac{(\Delta+\delta-\tau)^{\alpha-1}}{\Gamma(\alpha)} D_{f_2} K^\beta(\tau)d\tau \quad . \quad (50)$$
$$= -\frac{D_{f_2}(\gamma g)^\beta}{\Gamma(1+\alpha)}\left\{\alpha(\Delta+\delta)^{\alpha+\beta}\left[B(\beta+1,\alpha) - B\left(\frac{\delta}{\Delta+\delta};\beta+1,\alpha\right)\right] + \delta^\beta(\Delta^\alpha - \delta^\alpha) + \frac{\alpha\delta^{\alpha+\beta}}{(\alpha+\beta)}\right\}$$

In the literature [16], the ISA method gives an approximation PFG signal attenuation expression, $S(t) = E_{\alpha,1}\left(-\int_0^t DK^\beta(t')dt'^\alpha\right)$. If it is expanded and only the first and second terms of the expansion is kept, we obtain

$$S(t) = E_{\alpha,1}\left(-\int_0^t DK^\beta(t')dt'^\alpha\right)$$
$$\approx 1 - \frac{D_{f_2}(\gamma g)^\beta}{\Gamma(1+\alpha)}\left\{\alpha(\Delta+\delta)^{\alpha+\beta}\left[B(\alpha,\beta+1) - B\left(\frac{\Delta}{\Delta+\delta};\alpha,\beta+1\right)\right] + \delta^\beta(\Delta^\alpha - \delta^\alpha) + \frac{\alpha\delta^{\alpha+\beta}}{(\alpha+\beta)}\right\},$$

(51), the same as that given by Eqs. (49)-(50) (Note $B(x,y) = B(y,x)$ and $B(a;x,y) = B(1-a;y,x)$). The agreement between the result here and that obtained by ISA method can be explained by the following: in PGSE or PGSTE experiments as shown in Fig. 1, $K(\Delta+\delta-\tau') = K(\tau')$ creating

$$-\int_0^{\Delta+\delta}\frac{(\Delta+\delta-\tau)^{\alpha-1}}{\Gamma(\alpha)}D_{f_2}K^\beta(\tau)d\tau \xrightarrow{\tau'=\Delta+\delta-\tau} = -\int_0^{\Delta+\delta}\frac{\tau'^{\alpha-1}}{\Gamma(\alpha)}D_{f_2}K^\beta(\Delta+\delta-\tau')d\tau'. \qquad (52)$$

Eq. (52) is true in the typical PFG experiments where $|\gamma g_1\delta_1| = |\gamma g_2\delta_2|, |g_1| = |g_2|, \delta_1 = \delta_2$. The approximation expression, $S(t) = E_{\alpha,1}\left(-\int_0^t DK^\beta(t')dt'^\alpha\right)$ will deviate slightly from the solution here when the higher order terms are to be considered. In addition, Eq. (52) explains that at small attenuation, the results between fractal and fractional derivative model are equivalent. For time-fractional diffusion, $\beta = 2$, Eq. (49) can be further written as

$$S_1(\Delta+\delta) = -\frac{D_{f_2}(\gamma g)^\beta}{\Gamma(1+\alpha)} \times$$
$$\left\{\frac{2}{(\alpha+1)(\alpha+2)}\left[(\Delta+\delta)^{2+\alpha} - \Delta^{2+\alpha}\right] - \frac{2}{(\alpha+1)}\Delta^{1+\alpha}\delta - \Delta^\alpha\delta^2\right\} + \delta^\beta(\Delta^\alpha - \delta^\alpha) + \frac{\alpha\delta^{\alpha+\beta}}{(\alpha+\beta)}\right\}. \qquad (53)$$

Eq. (39) includes the $T_2$ relaxation in anomalous diffusion. The solution of Eq. (39) is

$$S(t) = \sum_{n=0}^{\infty} S_n(t), \qquad (54a)$$

where

$$S_0(t) = \sum_{k=0}^{m-1} S^{(k)}(0^+)\frac{t^k}{k!}, \, m-1 < \alpha < m, \qquad (54b)$$

$$S_n(t) = \int_0^t\left\{\left(\frac{(t-\tau)^{\alpha-1}}{\Gamma(\alpha)}D_fK^\beta(t) - \frac{1}{T_2}\right)S_{n-1}(\tau)\right\}d\tau. \qquad (54c)$$

It is hard to integrate the higher order terms in Eqs. (54a) and (54b) and (54c), as that requires computer assistance.

## VI. SIMULATION: CTRW ON LATTICE

Simulations were performed to verify the general solution, Eq. (46). The same PFG anomalous diffusion simulation method has been employed in [16]. The simulation is based on two models. The CTRW models [36] and the Lattice model [47, 48]. The CTRW model has been developed by [36], which is used to produce the waiting time and jump length in the simulation. The Lattice model developed by Ediger et al. was modified to record the $\phi_i(t) = \sum_{j=1}^{n} \gamma g(t_j) z(t_j) \tau_j$ [16,48]. The PFG signal attenuation in the simulation is obtained by $A(q,t) = \frac{1}{N_{walks}} \sum_{i=1}^{N_{walks}} \cos[\phi_i(t)]$ [16,48]. A total of 100000 walks were performed for each simulation. As the CTRW model [36] is proposed only for the subdiffusion simulation and is based on the fractional derivative model, the simulation results here is limited to the subdiffusion based on the fractional derivative. The readers are referred to get more detailed information in Refs. [36,16,47,48].

## VII. RESULTS AND DISCUSSION

TABLE I. Compared the results obtained in this paper with other reported results ($S(0)$ is set as 0).

| |
|---|
| Fractal derivative |
| Differential modified Bloch equation: $\frac{\partial M_{xy}(z,t)}{\partial t} = \alpha t^{\alpha-1} D_{f_1} \frac{\partial}{\partial z^\beta} M_{xy}(z,t) - i\gamma \mathbf{g}(t) \cdot z M_{xy}(z,t) - \frac{M_{xy}(z,t)}{T_2}$ |
| Homogeneous sample $M_{xy}(z,t) = S(t)\exp(-i\mathbf{K}(t) \cdot z)$: |
| $\frac{\partial S(t)}{\partial t} = -\left(\alpha t^{\alpha-1} D_{f_1} K^\beta(t) + \frac{t}{T_2}\right) S(t)$  Solution: $S(t) = \exp(-\frac{t}{T_2}) \exp\left(-\int_0^t D_{f_1} K^\beta(t') dt'^\alpha\right)$ |
| Results in literature:  a) $\frac{\partial S(t)}{\partial t} = -Dg^2 t^{2-\nu} S(t), \nu = 1 - \alpha$ [a,*]   b) $S(t) = \exp\left(-\int_0^t D_{f_1} K^\beta(t') dt'^\alpha\right)$ [b] |
| Fractional derivative |
| Integral modified Bloch equation: |
| $M_{xy}(z,t) = \sum_{k=0}^{m-1} M_{xy}^{(k)}(z,0^+) \frac{t^k}{k!} + \int_0^t \left\{ \left( \frac{(t-\tau)^{\alpha-1}}{\Gamma(\alpha)} D_{f_2} \frac{\partial^\beta}{\partial|z|^\beta} - i\gamma \mathbf{g}(t) \cdot z - \frac{1}{T_2} \right) M_{xy}(z,t) \right\} d\tau$ |
| Homogeneous sample $M_{xy}(z,t) = S(t)\exp(-i\mathbf{K}(t) \cdot z)$: |
| $S(t) = \sum_{k=0}^{m-1} S^{(k)}(0) \frac{t^k}{k!} + \int_0^t \left\{ \left( \frac{(t-\tau)^{\alpha-1}}{\Gamma(\alpha)} D_f K^\beta(t) - \frac{1}{T_2} \right) S(\tau) \right\} d\tau$ |
| Neglect $T_2$ relaxation: $\quad _t D_*^\alpha S(t) = -K^\beta(t) D_{f_2} S(t)$, |
| Solution: $\quad S(t) = \sum_{n=0}^{\infty} S_n(t)$, where $S_0(t) = \sum_{k=0}^{m-1} S^{(k)}(0^+) \frac{t^k}{k!}, m-1 < \alpha < m$, and |
| $S_n(t) = J^\alpha(a(t) S_{n-1}(t)) = -\int_0^t \frac{(t-\tau)^{\alpha-1}}{\Gamma(\alpha)} D_{f_2} K^\beta(\tau) S_{n-1}(\tau) d\tau$ |

Single pulse attenuation: $S(t) = E_{\alpha,1+\beta/\alpha,\beta/\alpha}\left(-D_f(\gamma g)^\beta t^{\alpha+\beta}\right)$

Under SGP approximation: $S(\Delta) = E_{\alpha,1}\left(-D_f K_{SGP}^\beta \Delta^\alpha\right), K_{SGP} = \gamma g \delta$

At small attenuation: $S(t) \approx E_{\alpha,1}\left(-\int_0^t DK^\beta(t')dt'^\alpha\right)$

Results in literature:

a) Modified Bloch equation by Hangya et al. [c*]

$(D + i\gamma b(t) + 1/T_2)^\beta M_{xy}(\mathbf{r},t) = QM_{xy}(\mathbf{r},t)$

Homogeneous sample: ${}_t D_*^\alpha S(t) = -at^\beta S(t)$ and $S(t) = E_{\beta,1+\alpha/\beta,\alpha/\beta}(-\gamma^\alpha t^{\alpha+\beta} \int_g |\mathbf{g}\cdot\mathbf{y}|^\alpha m(d\mathbf{y}))$ [c,*]

b) Modified Bloch equation by Magin et al. [d]

$C_s \cdot {}_t D_*^\alpha M_{xy}(\mathbf{r},t) = -i\gamma(\mathbf{r}\cdot\mathbf{g}(t))M_{xy}(\mathbf{r},t) + C_s D_f \nabla^\beta M_{xy}(\mathbf{r},t)$, $C_t = \tau^{\alpha-1}$ and $C_s = \mu^{\beta-1}$

Homogeneous sample: [d]

$E_{\alpha,1}\left[-i\gamma G_z \tau(t/\tau)^\alpha\right]\exp\left[-B(t/\tau)^\alpha\right], B = \dfrac{2\Gamma(2-\alpha)D\gamma^2 G_z^2 \tau^3}{3\alpha^2 \Gamma(2-\alpha)}$

c) $S(t) \approx E_{\alpha,1}\left(-\int_0^t DK^\beta(t')dt'^\alpha\right)$ [b], $S(\Delta) = E_{\alpha,1}\left(-D_f K_{SGP}^\beta \Delta^\alpha\right), K_{SGP} = \gamma g \delta$ [b]

a. Reference [40], b. Reference [14,16], c. Reference [33], d. Reference [14]

* These literature results treat only a single gradient pulse attenuation. The typical PFG experiment usually has three periods including two gradient pulses and a delay time.

    Two types of modified Bloch equations were built for fractional diffusion. One is the fractional differential equation based on the fractal derivative, and the other is the fractional integral equation based on the fractional derivative. The general PFG signal attenuation expressions for fractional diffusion were derived from both types of modified equations for free fractional diffusion as well as two other different methods, one is the observing only signal intensity at the origin method proposed in this paper, and the other is the EPSDE method. The major theoretical formalism obtained in this paper were compared with other reported results and summarized in Table I. From the fractional differential equation based on the fractal derivative, the obtained PFG signal attenuation is $S(t) = \exp\left(-\int_0^t D_{f_1} K^\beta(t')dt'^\alpha\right)$, which exactly reproduces the results from EPSDE method [15] and ISA method [16] and is consistent with other reported results [49]. While from the fractional integral modified Bloch equation based on the fractional derivative model, the obtained PFG signal attenuation expression is Eq. (46), which is closely related to Mittag-Leffler Function and it can give the SGP approximation result Eq. (47), $S(t) = E_{\alpha,1}\left(-D_f(t)K_{SGP}^\beta t^\alpha\right)$, which reproduces results in the literature [15,16].

    As shown in Fig. 2, at small attenuations, the stretched exponential attenuation (SEF) and the Mittag-Leffler function (MLF) attenuation agree with each other, while at large attenuations, the stretched exponential function attenuates faster than that of the Mittag-Leffler function at subdiffusion (however, its attenuation is slower at superdiffusion as pointed out in literature [16]). The difference may be because the fractal derivative is a local operator while the fractional derivative is a global operator [34,35]. The different

attenuation speed can be seen more clearly under SGP approximation where MLF has $\frac{(t_3-\tau)^{\alpha-1}}{\Gamma(\alpha)}$ in Eq. (46c) while SEF has $\alpha t^{\alpha-1}$ in Eq. (21). When $D_{f_1} = D_{f_2}/\Gamma(1+\alpha)$, for subdiffusion the value of $\int_0^t D_{f_1} K^\beta(\tau)\alpha\tau^{\alpha-1}d\tau$ will be bigger than the value of $\int_0^t D_{f_2}(t_3-\tau)^{\alpha-1}K^\beta(\tau)d\tau/\Gamma(\alpha)$ as the large $K(\tau)$ showing up in the middle of the time domain as that described in Eq. (10). While for superdiffusion, the stretched exponential attenuation is slower than the Mittag-Leffler function attenuation. For subdiffusion, the Mittag-Leffler functional attenuation based on fractional derivative may be a favorite attenuation expression as it agrees with the PFG curvilinear diffusion. It has been theoretically derived that for penetrants diffusing along an ideal polymer chain in free or restricted domains, the PFG signal attenuation expression is a Mittag-Leffler function with $\alpha = 0.5$, in either isotropic or anisotropic diffusion [50,51]. For superdiffusion, the Mittag-Leffler function attenuation can give a negative value that remains unclear, which may be due to a fat tail in the probability function or to the initial condition $\frac{\partial}{\partial t}M_{xy}(z,0^+) = 0$ that needs to be adjusted, which requires further investigation.

The signal attenuation expressions also agree with other various methods. The SEF attenuation agrees reasonably with Karger's results and modified Gaussian SGP method. It is found that Wisdom et al.'s results based on spectral function method agree with the SEF attenuation during the first gradient pulse. The signal intensity equation proposed in Wisdom et al.'s paper only considered the signal attenuation in the first gradient pulse. The MLF attenuation agrees with the non-Gaussian approximation method. The CTRW simulation agrees perfectly with the MLF signal attenuation.

The fractional differential modified Bloch equation based on the fractal derivate has no corresponding reports yet. The fractional integral modified Bloch equation based on the fractional derivative model is a new type of modified Bloch equation. From Eqs. (36) and (38), the spin fractional diffusion may be referred to as a nonlinear process with respect to time while the precession and relaxation may be referred to as a linear process with respect to time. Inside the integral of Eq. (38), the contribution from the nonlinear process $\frac{(t-\tau)^{\alpha-1}}{\Gamma(\alpha)}D_f \frac{\partial^\beta}{\partial|z|^\beta}M(z,\tau)d\tau$ is combined with the linear process contribution, $(-i\gamma g(\tau)z - 1/T_2)M_{xy}(z,\tau)d\tau$, which is a real-time combination of different contributions at each interval during diffusion. For space-fractional diffusion, $\alpha = 1$, the nonlinear process reduces to a linear process, and the integral modified Bloch equation Eq. (38) reduces to Eq. (41), which agree with the modified Bloch equations based on the fractional derivative proposed by two groups [14,33]. The results of time-fractional diffusion from this paper are different from that in Ref. [14]. However, the results for general fractional diffusion is consistent with the result in Ref. [33] when PFG signal attenuation results from a signal gradient pulse. The Ref. [33] results only give the PFG signal attenuation for this single gradient pulse. The combination strategies adopted here are different from these strategies adopted in Refs. [14,33] and in building the modified Bloch equations. The modified Bloch equation proposed in [14] can be described as

$$C_s \cdot {}_t D_*^\alpha M_{xy}(\boldsymbol{r},t) = -i\gamma(\boldsymbol{r}\cdot \boldsymbol{g}(t))M_{xy}(\boldsymbol{r},t) + C_s D_f \nabla^\beta M_{xy}(\boldsymbol{r},t), \quad (55)$$

where $C_t = \tau^{\alpha-1}$ and $C_s = \mu^{\beta-1}$ are the fractional order time and space constants to preserve the units. Eq. (55) is a direct combination of different motion equations, which is the conventional way used in normal diffusion. The obtained PFG signal attention expression from Eq. (55) are $E_{\alpha,1}[-i\gamma G_z \tau (t/\tau)^\alpha]\exp[-B(t/\tau)^\alpha]$, $B = \dfrac{2\Gamma(2-\alpha)D\gamma^2 G_z^2 \tau^3}{3\alpha^2 \Gamma(2-\alpha)}$ which is not reducible to Eq. (47). If Eq. (55) is written in integral form, the combination inside the integral is

$$\frac{(t-\tau)^{\alpha-1}}{\Gamma(\alpha)} D_f \frac{\partial^\beta}{\partial |z|^\beta} M(z,\tau)d\tau + \frac{(t-\tau)^{\alpha-1}}{\Gamma(\alpha)}(-i\gamma g(\tau)z)M_{xy}(z,\tau)d\tau,$$ where the contribution from the linear

precession process is changed to a nonlinear contribution by multiplying $\dfrac{(t-\tau)^{\alpha-1}}{\Gamma(\alpha)}$. When the diffusion coefficient $D_f = 0$, this combination will result in an $M_{xy}(z,t) = S(t)E_{\alpha,1}(-i\gamma gzt)$ if a constant gradient is present. This magnetization is different from Eq. (11), $M_{xy}(z,t) = S(t)\exp(-i\boldsymbol{K}_{SGP,t}\cdot\boldsymbol{z})$ for non-diffusing spins. Another group proposed a different modified Bloch equation which can be described as

$$(D + i\gamma b(t) + 1/T_2)^\beta M_{xy}(\boldsymbol{r},t) = Q M_{xy}(\boldsymbol{r},t), \quad (56a)$$

where

$$(D+a)^\beta f(t) := e^{-at} D^\beta [e^{at} f(t)]$$
$$= \exp(-i\gamma \boldsymbol{r}\cdot\boldsymbol{g}(t)t)\frac{1}{\Gamma(m-\alpha)}\int_0^t \frac{\frac{d^m}{d\tau^m}[\exp(-i\gamma \boldsymbol{r}\cdot\boldsymbol{g}(\tau)\tau)f(\tau)]}{(t-\tau)^{\alpha+1-m}} d\tau, m-1 < \alpha < m \quad (56b)$$

$b(t) = \boldsymbol{r}\cdot\boldsymbol{g}(t)$, $D^\beta$ denotes Caputo derivative, and $Q$ denotes the $\nabla^\beta$ that is equal to space fractional derivative defined in Appendix B. The combination in Eq. (56) may not easily to be applied to non-constant gradient fields as the $\exp(-i\gamma\boldsymbol{r}\cdot\boldsymbol{g}(t)t)$ before the integral and $\exp(-i\gamma\boldsymbol{r}\cdot\boldsymbol{g}(\tau)\tau)$ inside the integral will be different. The practical gradient pulse can be a shaped pulse such as a Gaussian shaped pulse, which has a benefit to increasing or decreasing the gradient intensity smoothly. There are no PFG signal attenuation results for the periods from $\delta < t \le \Delta + \delta$ reported in ref. [33] for time fractional diffusion and general fractional diffusion. The term $\exp((i\gamma\boldsymbol{r}\cdot\boldsymbol{g}(t) - 1/T_2)t)$ in Eq. (56) may be used to counteract the $\exp(-(i\gamma\boldsymbol{r}\cdot\boldsymbol{g}(t) - 1/T_2)t)$ of the magnetization to leave only signal intensity terms on both sides of the diffusion equation, which is equal to the observing only the signal intensity at the origin method as proposed in section [IV] for the first period in the PGSE experiment as shown in Fig. 1. If $\exp(i\boldsymbol{K}(t)\cdot\boldsymbol{r})$ can be used to replace $\exp((i\gamma \mathbf{r} \cdot \mathbf{g}(t))t)$, we can get $D^\beta[\exp(i\mathbf{K}(t) \cdot \mathbf{r})M_{xy}(\mathbf{r},t)] = \exp(i\mathbf{K}(t) \cdot \mathbf{r})QM_{xy}(\mathbf{r},t)$ which will give the same result for fractional diffusion in a homogeneous sample as Eqs. (22) and (28). However, in a system with heterogeneous initial distribution or restricted diffusion, the phase cannot be counteracted by $\exp((i\gamma \mathbf{r} \cdot \mathbf{g}(t) - 1/T_2)t)$ perfectly. Additionally, it may not be possible to assume that we still directly use $\exp(-t/T_2)$ as a solution of the modified Bloch equation based on the fractional derivative directly. From the numerical calculation, even in the homogeneous system, the relaxation behavior deviates from $\exp(-t/T_2)$ which will be discussed later. The solution in Ref. [33] is

$$S(t) = E_{\beta,1+\alpha/\beta,\alpha/\beta}(-\gamma^\alpha t^{\alpha+\beta} \int_g |\mathbf{g}\cdot\mathbf{y}|^\alpha m(d\mathbf{y})), \text{ where } E_{\beta,\alpha,\gamma} = \sum_{n=0}^{\infty}\sum_{k=0}^{n-1}\frac{\Gamma(\beta(\alpha k+\gamma)+1)}{\Gamma(\beta(\alpha k+\gamma+1)+1)}$$

, which is equal to Eq. (48) obtained in the first gradient pulse. The merit of the integral type modified Bloch equation is that it provides a new way to combine different evolution processes, where the original properties of the contributions from linear or nonlinear processes remain unchanged at the instant of the combination.

The modified Bloch equation can be understood in a process decoupling context: in PFG experiments, at each small interval from $t$ to $t+dt$, the spin magnetization undergoes a few simultaneous processes. We can separate these processes during $dt$ as consecutive processes without changing the outcome. Arbitrarily, the diffusion takes place first, followed by relaxation and precession consecutively. For the diffusion process, the wavenumber vector is kept constantly as $\mathbf{K}(t)$, which keeps its value the same as that at the starting point of the interval from $t$ to $t+dt$. This decoupled diffusion process will result in a signal attenuation $\frac{(t-\tau)^{\alpha-1}}{\Gamma(\alpha)}D_f\frac{\partial^\beta}{\partial|z|^\beta}M(z,\tau)d\tau$. The relaxation follows behind the diffusion will yield another signal attenuation $\frac{1}{T_2}M_{xy}(z,t)$. The precession behind the relaxation only changes the phase of the magnetization, which is the only process among these three processes that may be not visible by observers with limited instrument resources. The diffusion and relaxation will have an accumulating effect which can be recorded in an integral way by the integral equation Eq. (38).

The integral type of modified Bloch equation provides a good way to combine processes described by different time-derivatives such as $\partial/\partial t$ and ${}_tD_*^\alpha$. The differential type modified Bloch equation reported in literature can be converted to integral type equations. For instance, the Eq. (5) can be transferred to

$$M(z,t) = M(z,t) + \int_0^t \left( D_f \frac{\partial^2}{\partial z^2} - i\gamma g(\tau)z \right) M(z,\tau)d\tau, \qquad (57)$$

which is the integral type modified Bloch equation for normal diffusion, while, Eq. (30) can be transferred to

$$M_{xy} = M_{xy} + \int_0^t \left[ \alpha t^{\alpha-1} D_f \nabla^\beta M_{xy} - i\gamma g(t)zM_{xy} - \frac{M(z,t)}{T_2} \right] dt, \qquad (58)$$

which is the integral type modified Bloch equation for anomalous diffusion based on the fractal derivative. Eqs. (57) and (58) can be transformed back to to the differential equation easily by operating $\partial/\partial t$ on both sides of the equation. However, for the fractional derivative model, it is difficult to transfer the integral equation to a differential equation without resorting to approximation. The integral equation provides an alternative way to solve the PFG signal attenuation expression. For instance, the Adomian decomposition method can be used to solve the Eq. (57) for PFG normal diffusion. As an illustration, $\gamma g = 1$ is used and only the first gradient effect will be considered. The first six lower order terms are given in the following:

$S_0(t) = 1$, $S_1(t) = -izt$, $S_2(t) = (izt)^2/2$, $S_3(t) = -Dt^3/3 - (izt)^3/3!$, $S_4(t) = (izt)Dt^3/3 + (izt)^4/4!$,

$S_5(t) = -(izt)^2 Dt^3/3 - (izt)^5/5!$, and $S_6(t) = 1/2(Dt^3/3)^2 + (izt)^3 6Dt^3/3 + (izt)^6/6!$ ⋯. The above terms from the Adomian decomposition method reproduce the Tylor expansion of familiar result $S_3(t) = \exp(-Dt^3/3 - izt)$ for the PFG signal attenuation of normal diffusion during the first gradient. The above calculation procedure by the Adomian decomposition method is tedious. However, sometimes it is difficult to find other methods, then the Adomian decomposition method [41-45] could give a solution.

The numerical calculation for Eq. (46) for general fractional diffusion needs computer assistance. The CTRW simulations agree with the theoretical results as shown in Figs. 3-4. The diffusion constant $D_{f_2} = 1.85 \times 10^{-10}$ m$^\beta$/s$^\alpha$ obtained from fitting $\langle z^\beta(t) \rangle$ versus $t$ in Fig. 3 was used in the theoretical calculation of Eq. (46) in Figs. 4 (a)-(b). In Figs. 4 (a)-(b), the derivative order parameters are $\alpha = 0.6, \beta = 2$ and the $g$ equals 0.1 $T/m$ are used. Fig. 4 (a) shows that there is agreement between the theoretical results and CTRW simulations at various $\Delta - \delta$ values equaling 0 ms, 10 ms and 20 ms. Fig. 4 (b) shows the agreement between the theoretical result and CTRW simulation under SGP approximation, where $\gamma g \delta = 6 \times 2.675 \mathrm{k} \times 10^4$ was used. The numerical calculation is simple for small signal attenuation, but for large signal attenuation it may encounter overflow, which is a common issue in the special function evaluation. There are many efforts dedicated to the special function evaluation such as the Mittag-Leffler function [52]. As Eq. (46) is new, it still requires many mathematical efforts for improving the numerical calculation. Additionally, it also requires finding some alternative solution methods to solve Eq. (44) which may make the calculation much easier.

Fig. 5 compares the PFG signal attenuations obtained from the fractional derivative model with and without $T_2$ relaxation effect. These same diffusion conditions: $\alpha = 0.6, \beta = 2$, $D_{f_2} = 1.85 \times 10^{-10}$ m$^\beta$/s$^\alpha$, $\Delta - \delta = 0$ ms and $g$ equaling 0.1 $T/m$ are used in Fig. 5. $T_2$ is 50 ms. Fig. 5 indicates that the signal attenuation $S(t)$ based on Eq. (54) is not proportional to $\exp(-\frac{t}{T_2})$. From the deviation, the effective $T_2$ relaxation in PFG anomalous diffusion is slower than that without influence by anomalous diffusion. This is significantly different from the normal diffusion and the fractal derivative model. In both of them, $S(t)$ is proportional to $\exp(-\frac{t}{T_2})$. This deviation needs to be investigated by further experimental and theoretical efforts.

In practical application, when the signal attenuation is moderate and the equal width gradient pulses are employed in PGSE or PGSTE experiments, Eqs. (49)-(51) in Section [V.B.2] indicate that the signal attenuation is approximately equal to $S(t) = E_{\alpha,1}\left(-\int_0^t DK^{\beta}(t')dt'^{\alpha}\right)$. That has been obtained by the instantaneous signal attenuation approximation method [16]. Therefore, $S(t) = E_{\alpha,1}\left(-\int_0^t DK^{\beta}(t')dt'^{\alpha}\right)$ can be an alternative way to analyze the data if signal attenuation is moderate. Mittag-Leffler function can be calculated by the software in Matlab as well [53]. Additionally, the time and space fractional derivative orders $\alpha, \beta$ cannot be arbitrarily set during the fitting of PFG experimental data. Pointed out in reference [16], at small attenuation, the $\alpha, \beta$ can be determined by

$$\ln[\ln(S(0) - \ln S(t)] = \begin{cases} c_1 + \beta \ln(g), & \text{when } \delta \text{ and } \Delta \text{ are fixed} \\ c_2 + (\alpha + \beta) \ln(\delta), & \text{when } \Delta = \delta, \text{ and } g \text{ are fixed} \\ c_3 + \alpha \ln(\Delta), & \text{when } \delta \ll \Delta, \text{and } g \text{ are fixed} \end{cases}, \quad (59)$$

where $c_1$, $c_2$ and $c_3$ are constants. Please referred to [16] for more detailed information.

The magnetization of the spins at the same position is treated as an ensemble by the modified Bloch equation, while the EPSDE method treats each spin as an individual in the phase diffusion process. The observing only signal intensity method (or observing the signal intensity at the origin method) is simple, which provides an alternative way to solve PFG signal attenuation, that can give an exact solution only in free normal or anomalous diffusion in a homogeneous sample. The observing signal intensity at the origin method neglects the phase change by observing at the origin, while the modified Bloch Equation method considers the coeffct of the two processes at each interval. For a homogeneous spin system, all three methods give the same PFG signal attenuation result. Further efforts are still needed in this field for various situations such as restricted anomalous diffusion and anisotropic anomalous diffusion [50]. The new general modified Bloch equations and the general solution is essential to PFG anomalous diffusion in NMR and MRI research.

**Figure 1**

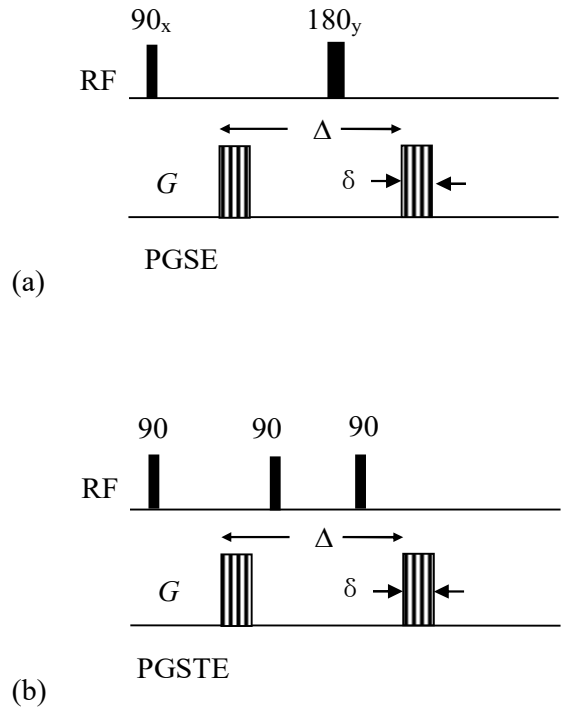

(a)

(b)

**Fig. 1** (a) PGSE pulse sequences, (b) PGSTE pulse sequence. The gradient pulse width is $\delta$, and the diffusion delay is $\Delta$.

**Figure 2**

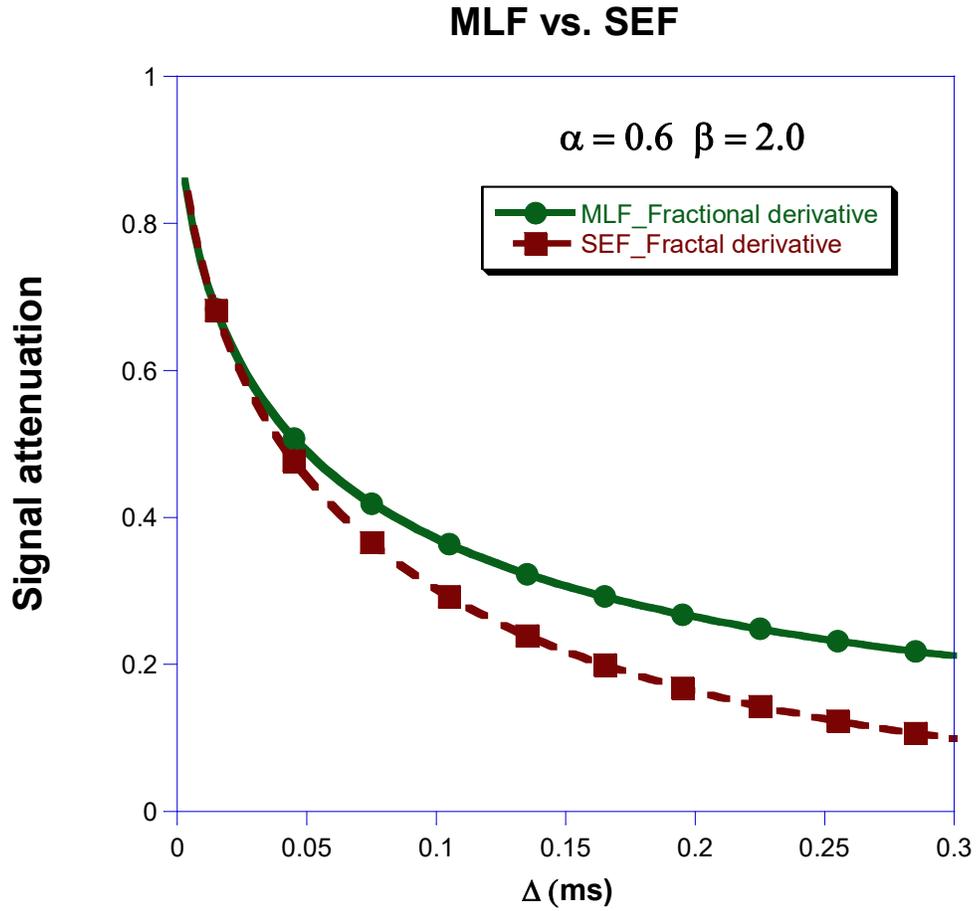

**Fig. 2** Comparison of the signal attenuation from the fractal derivative Eq. (21) with that from the fractional derivative Eq. (47). The signal attenuation results obtained by SGP approximation are used. The parameters used are $\alpha = 0.6, \beta = 2$, $D_{f_2} = D_{f_1}\Gamma(1+\alpha) = 1.85 \times 10^{-10}$ m$^\beta$/s$^\alpha$, and $\gamma g \delta = 6 \times 2.6751 \times 10^4$.

**Figure 3**

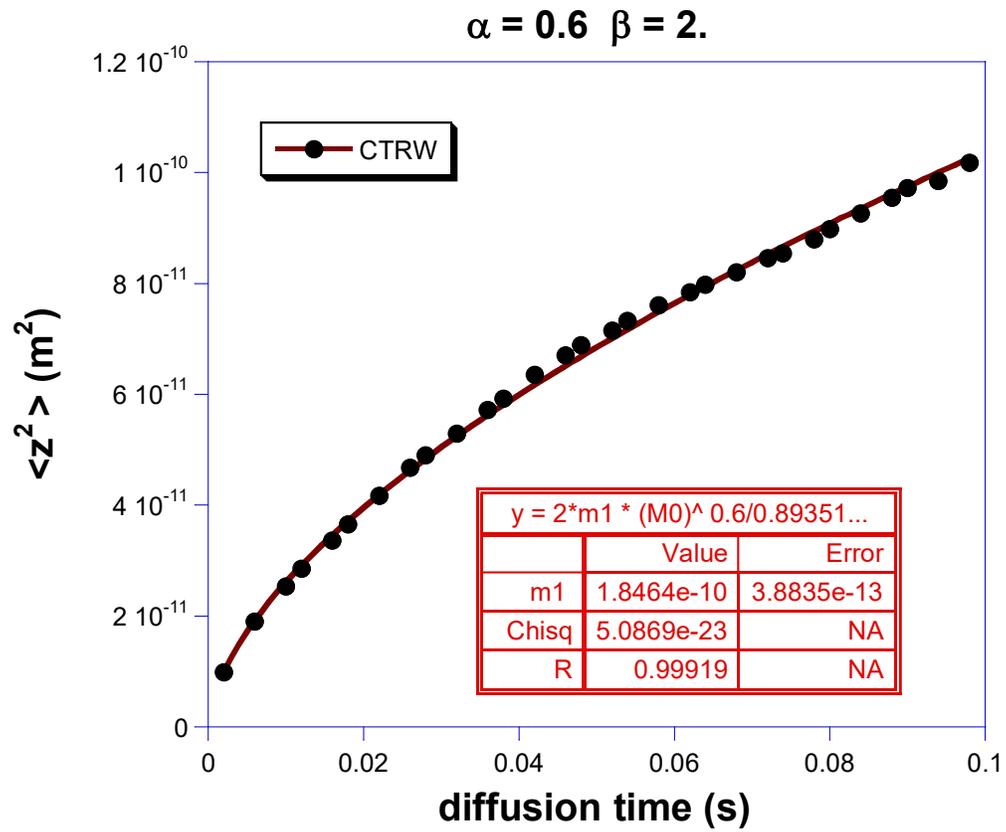

**Fig. 3**. $\langle z^\beta(t) \rangle$ versus $t$ from the simulation: (a) $\alpha = 0.6, \beta = 2$. The fractional diffusion constant determined from the fitting is $D_{f_2} = 1.85 \times 10^{-10}$ m$^\beta$/s$^\alpha$.

**Figure 4(a)**

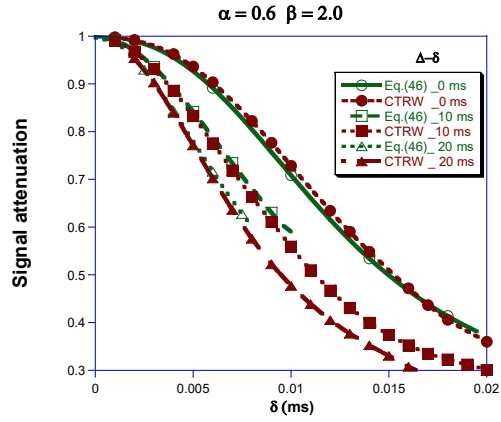

**Figure 4(b)**

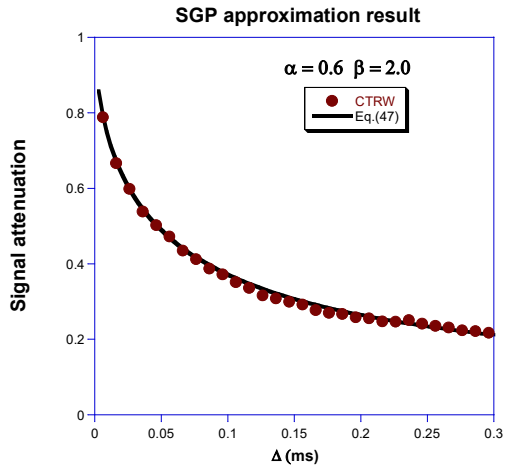

**Fig. 4** Comparison PFG signal attenuation from Eq. (46) with that obtained from CTRW simulation, $\alpha = 0.6, \beta = 2$ and $D_{f_2} = 1.85 \times 10^{-10}$ m$^\beta$/s$^\alpha$ : (a) finite gradient pulse width effect with $\Delta - \delta$ equaling 0 ms, 10 ms and 20 ms, $g$ equaling 0.1 $T/m$ , (b) SGP approximation result, $\alpha = 0.6, \beta = 2$ , $D_{f_2} = D_{f_1}\Gamma(1+\alpha) = 1.85 \times 10^{-10}$ m$^\beta$/s$^\alpha$, and $\gamma g \delta = 6 \times 2.6751 \times 10^4$ .

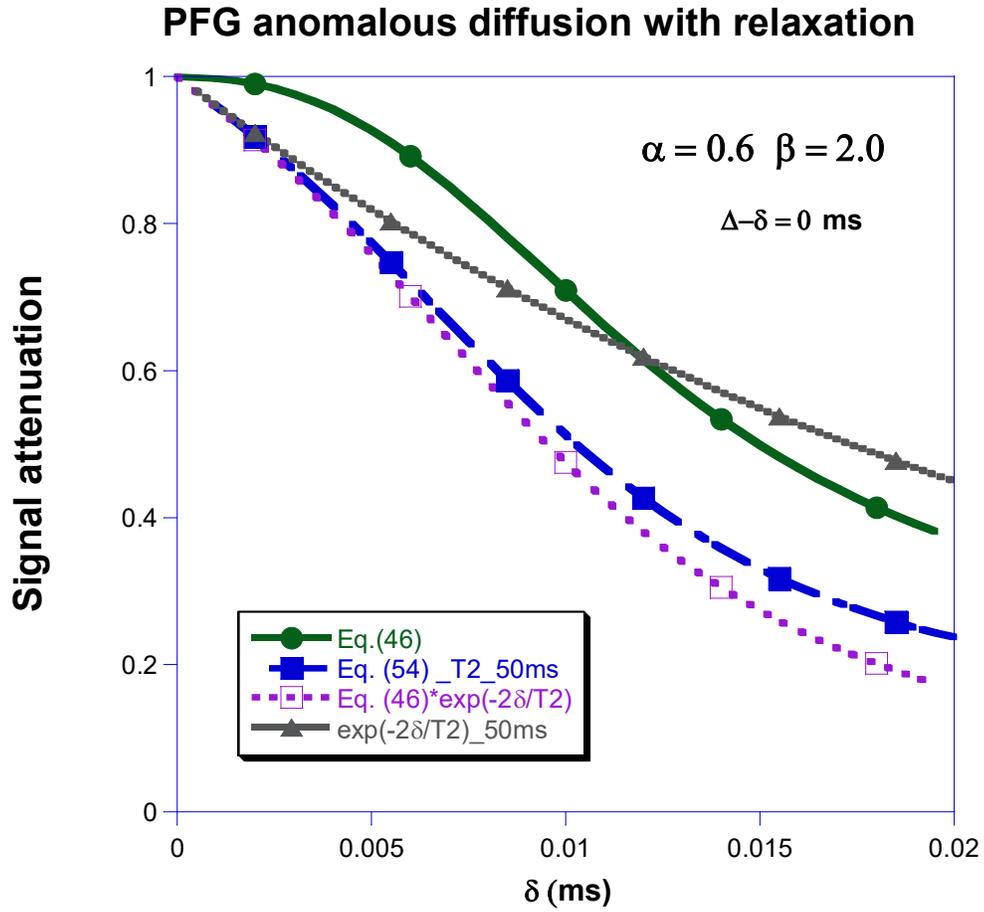

**Fig. 5** Spin-spin relaxation effect on PFG anomalous diffusion, as described by the fractional derivative. $\alpha = 0.6, \beta = 2$, $D_{f_2} = 1.85 \times 10^{-10}$ m$^\beta$/s$^\alpha$, $\Delta - \delta = 0$ ms, and $T_2$ is 50 ms are used. The signal attenuation $S(t)$ is not proportional to $\exp(-\frac{t}{T_2})$.

## ACKNOWLEDGEMENTS

The linguistic help from my colleague Frank Abell, the students Ashley Burke, and Alex Wall from chemistry department, Thomas Caywood and Emma-Rose Sonberg from writing center, and Amoy Lin is acknowledged.

## APPENDIX A: DEFINITION OF THE FRACTAL DERIVATIVE.

From reference [11-12], the definition of the fractal derivative is

$$\frac{\partial P^{\beta/2}}{\partial t^{\alpha}} = \lim_{t_1 \to t} \frac{P^{\beta/2}(t_1) - P^{\beta/2}(t)}{t_1^{\alpha} - t^{\alpha}}, 0 < \alpha, 0 < \beta/2. \tag{A.1}$$

## APPENDIX B: DEFINITION OF THE FRACTIONAL DERIVATIVE.

The definition of the space fractional derivative [3,13-15] is given by

$$\nabla^{\beta} = \frac{\partial^{\beta}}{\partial |x|^{\beta}} + \frac{\partial^{\beta}}{\partial |y|^{\beta}} + \frac{\partial^{\beta}}{\partial |z|^{\beta}} \quad \frac{\partial^{\beta_{x'}}}{\partial |x'|^{\beta_{x'}}}$$

$$\frac{d^{\beta}}{d|z|^{\beta}} = -\frac{1}{2\cos\frac{\pi\alpha}{2}} \left[ {}_{-\infty}\boldsymbol{D}_z^{\beta} + {}_z\boldsymbol{D}_\infty^{\beta} \right],$$

where

$$_{-\infty}\boldsymbol{D}_z^{\beta} f(z) = \frac{1}{\Gamma(m-\beta)} \frac{d^m}{dz^m} \int_{-\infty}^{z} \frac{f(y)dy}{(z-y)^{\beta+1-m}}, \beta > 0, m-1 < \beta < m, \tag{B.1}$$

and

$$_z\boldsymbol{D}_\infty^{\beta} f(z) = \frac{(-1)^m}{\Gamma(m-\beta)} \frac{d^m}{dz^m} \int_{z}^{\infty} \frac{f(y)dy}{(y-z)^{\beta+1-m}}, \beta > 0, m-1 < \beta < m. \tag{B.2}$$